\documentclass{webofc}
\usepackage[varg]{txfonts}   % Web of Conferences font
\usepackage{hyperref}
\usepackage{url}
\usepackage{amsmath,amssymb,bm}
\usepackage{gensymb}
\usepackage{float}
\hypersetup{colorlinks=true,citecolor=blue,urlcolor=blue,linkcolor=blue}
\usepackage{array,multirow,graphicx}
\begin{document}
\title{Large-scale anisotropies
of ultra-high-energy cosmic rays measured
at the Pierre Auger Observatory
}

\author{\firstname{Marta} \lastname{Bianciotto}\inst{1}\fnsep\thanks{\email{spokespersons@auger.org}} for the Pierre Auger Collaboration\inst{2}
}

\institute{Università degli studi di Torino \& INFN, Sezione di Torino, Via Giuria 1, Torino, Italy 
\and
           Observatorio Pierre Auger, Av. San Martín Norte 304, 5613 Malargüe, Argentina \\ Full author list: \href{https://www.auger.org/archive/authors_2024_09.html}{https://www.auger.org/archive/authors\_2024\_09.html}
          }

\abstract{%
  Measurements of anisotropies in the arrival directions of ultra-high-energy cosmic rays are crucial to pinpoint their sources, which are yet to be discovered.
A dipolar anisotropy in right ascension above 8~EeV has been detected by the Pierre Auger Observatory with a significance of $6.8 \sigma$. The direction of the dipole suggests an extragalactic origin of ultra-high-energy cosmic rays above those energies.
In this contribution, we provide an overview of the studies on large-scale anisotropies in the arrival directions of ultra-high-energy cosmic rays measured at the Pierre Auger Observatory with energy thresholds from $\sim 0.03$~EeV up to $32$~EeV and we present and discuss the recent results achieved with the latest available dataset, which includes 19 years of operations -- resulting in a total exposure of 123,000~km$^2$~sr~yr and nearly 50,000 events above 8~EeV.
}
\maketitle
\section{Introduction}
\label{intro}
Ultra-high energy cosmic rays (UHECRs) are generally defined as charged astroparticles with energies exceeding $10^{18}$~eV, whose energy spectrum, mass composition, and arrival direction distribution can be measured experimentally via ground-based detectors \cite{general}.
A central question in high-energy astrophysics concerns the origin of UHECRs: during their propagation, these charged particles are deflected by extragalactic and Galactic magnetic fields, making their correlation with possible sources non-trivial to determine.
Significant progress in the search for UHECR sources has been made thanks to the data from the Pierre Auger Observatory \cite{PierreAuger:2015eyc}, the world’s largest area and exposure detector for UHECRs, which covers an area of $\sim 3000$~km$^2$. % with more than 1600 water-Cherenkov detectors and 27 fluorescence telescopes.
The Observatory is based on a hybrid detection technique, combining a surface detector array that measures secondary particles at the ground level and fluorescence detector telescopes that measure the development of the air showers in the atmosphere.

In this contribution, we present an overview of the studies on large-scale anisotropies in the arrival directions of UHECRs measured at the Pierre Auger Observatory with energy thresholds from $\sim$ 0.03~EeV up to 32~EeV based on \cite{PierreAuger:2024fgl} and we discuss the results. These studies are based on 19 years of data from the Observatory, resulting in a total exposure of 123,000~km$^2$~sr~yr and nearly 50,000 events above 8~EeV. 
In particular, we present the study of the dipole modulation in right ascension (R.A., or $\alpha$) and azimuth ($\phi$) above 4~EeV, also in the case that a quadrupolar distribution is present, as well as of the dipole modulation in R.A. down to 0.03~EeV. Finally, we provide an update on the angular power spectrum above 4~EeV.

\section{Analyses and results}
\label{sec-1}

\subsection{3D dipole above 4~EeV}
\label{subsec-1}

\begin{figure}[h]
\centering
\includegraphics[width=5cm,clip]{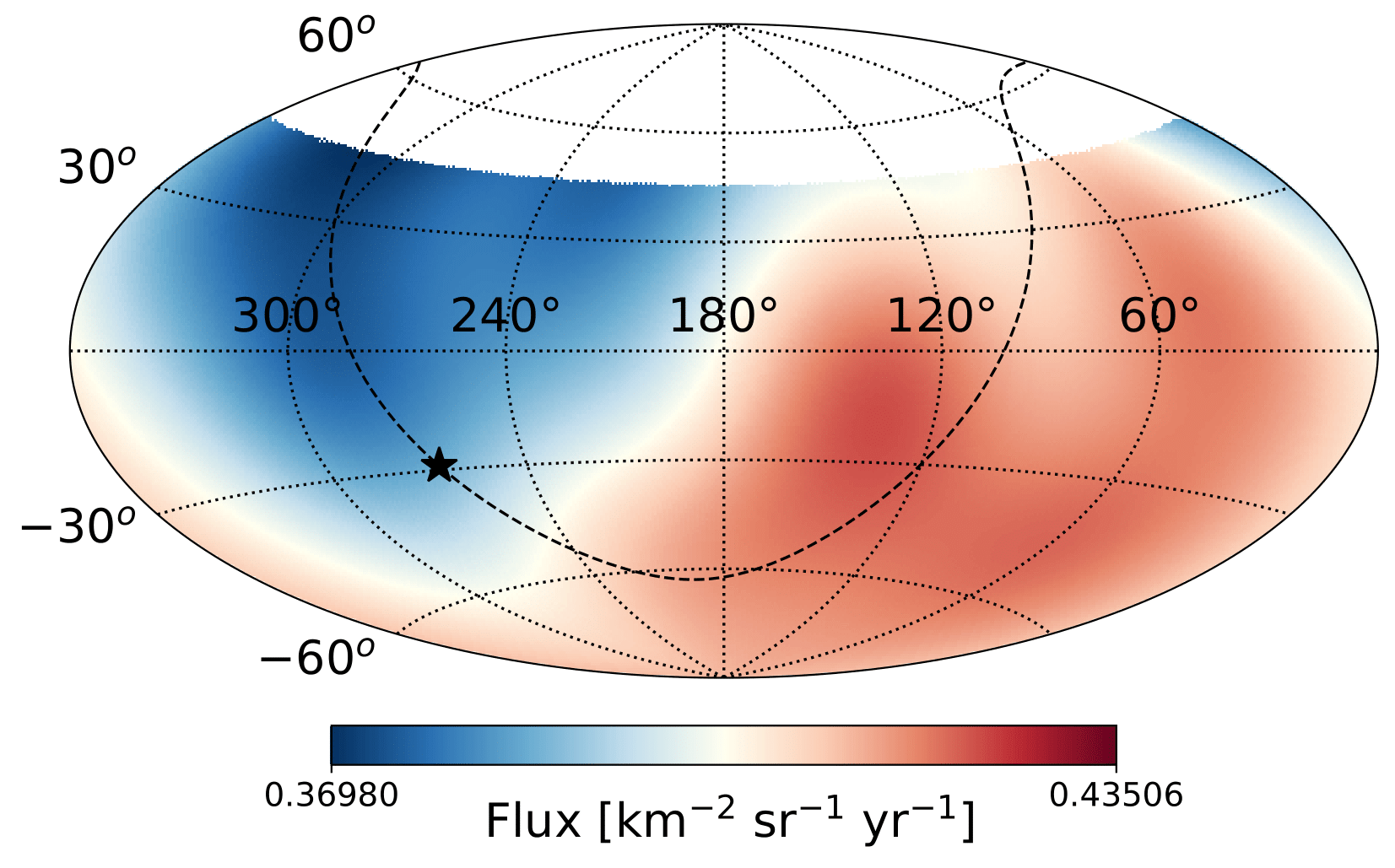}
\includegraphics[width=5cm,clip]{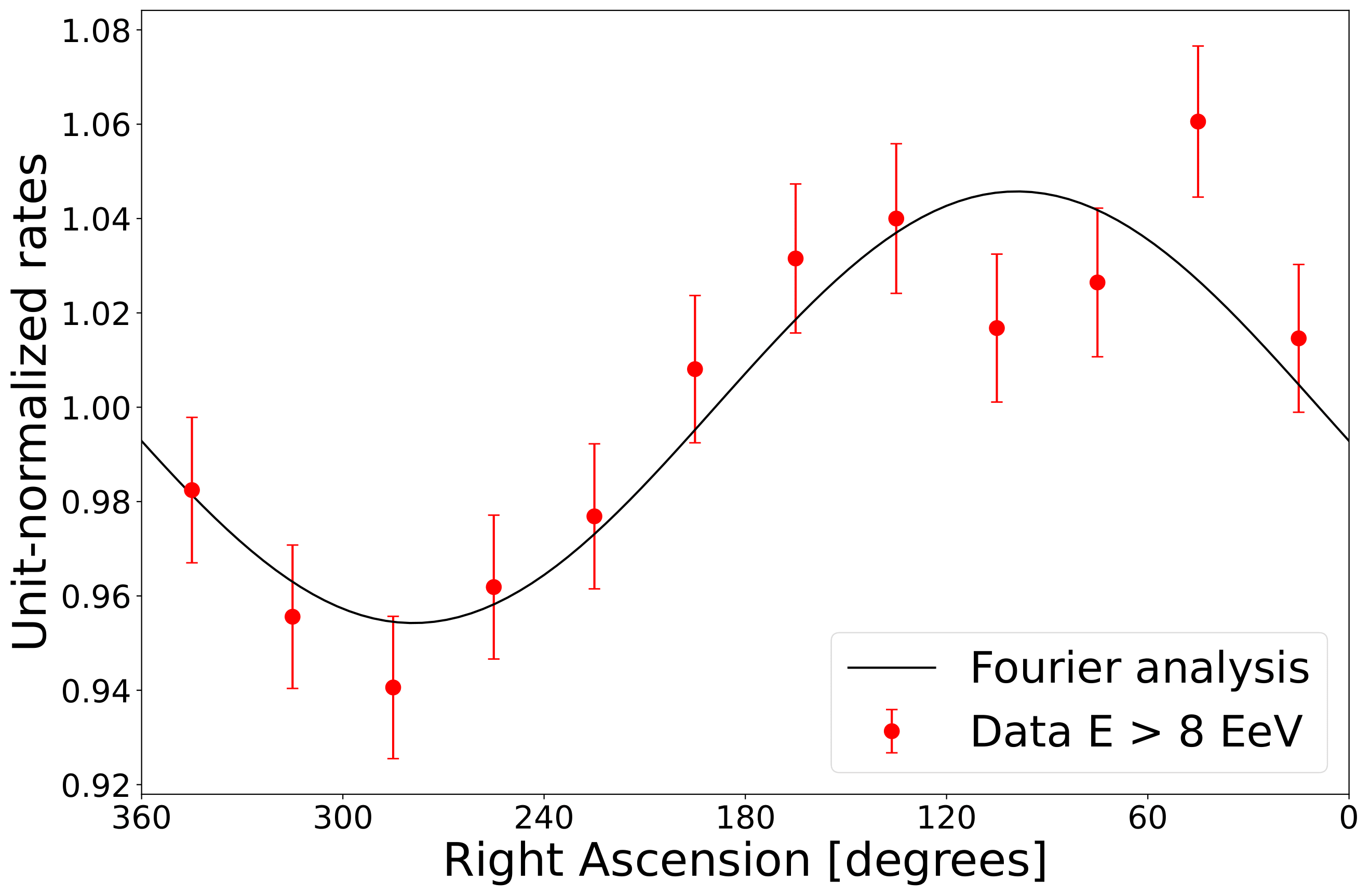}
\caption{Left: Flux above 8~EeV, smoothed by a Fisher distribution 
with a mean cosine of the angular distance to the center of the window equal to that of a top-hat distribution with radius of $45\degree$, in Equatorial coordinates. The Galactic center is represented by a black star and the Galactic Plane is represented by a black dashed line. Right: Normalized rates distribution in R.A. (red dots) with the predicted modulation obtained from the Rayleigh analysis (black line).}
\label{fig-1}
\end{figure}

At energies where the Observatory operates at full efficiency, a separate Fourier analysis in R.A. and azimuth can be conducted to make a 3D reconstruction of the dipole \cite{Aab_2015}. The analysis in R.A. is sensitive to the equatorial ($d_\perp$) component of the dipole anisotropy, while the analysis in azimuth is sensitive to the North-South ($d_z$) component. The analysis was done for four energy bins, (4-8, 8-16, 16-32, $\geq$ 32)~EeV, as well as for a cumulative bin above 8~EeV\@.
In Fig.~\ref{fig-1}, the flux above 8~EeV, in equatorial coordinates, and the dipole modulation in R.A. are shown. For $E \geq 8$~EeV, the significance is now at $6.8 \sigma$, while for the 8-16~EeV bin is $5.7\sigma$, against the isotropic null hypothesis. The direction of the dipole is pointing $115~\degree$ away from the Galactic center, suggesting the extragalactic origin of cosmic rays above those energies.

\begin{figure}[h]
\centering
\includegraphics[width=6cm,clip]{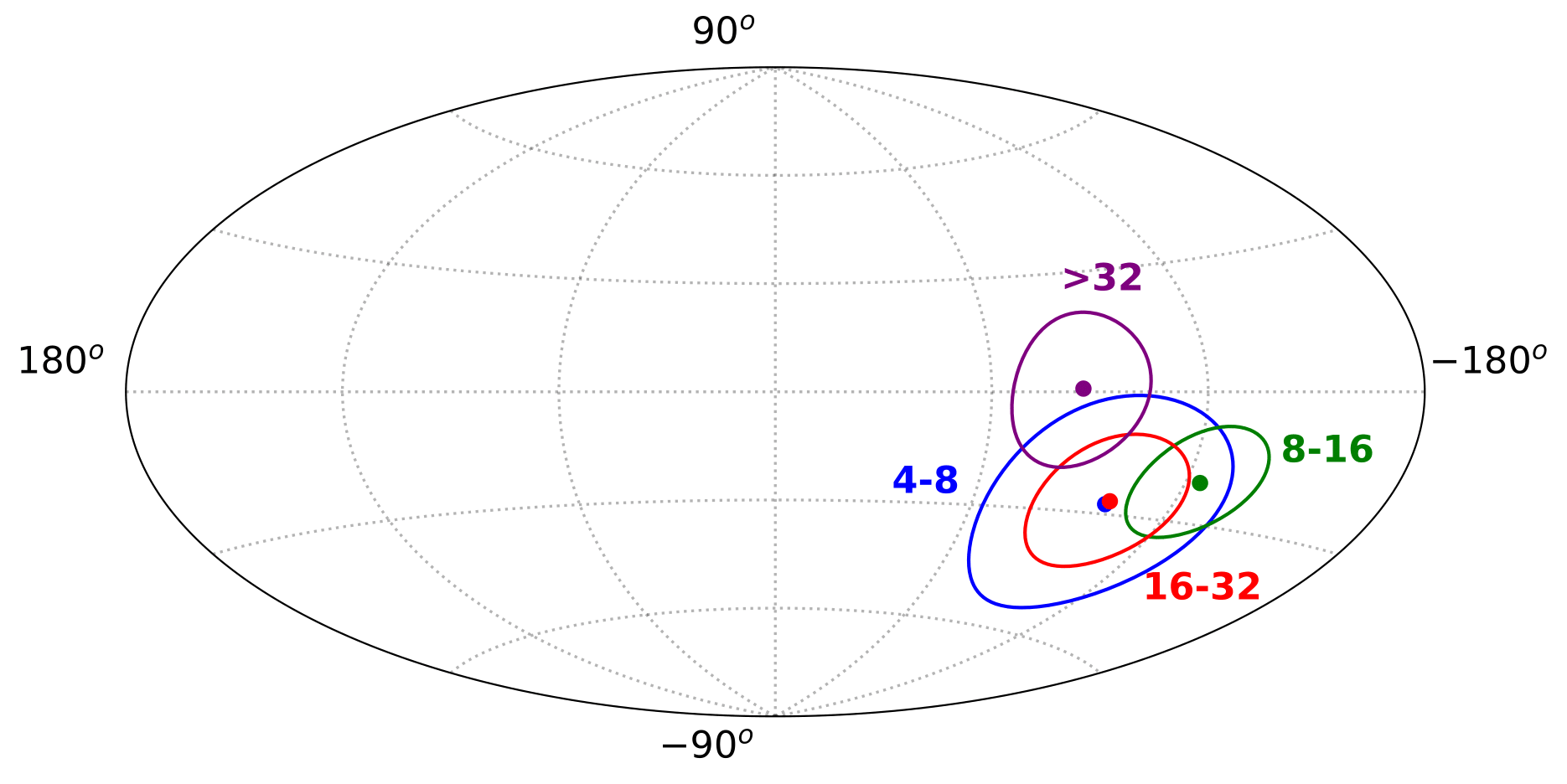}
\includegraphics[width=5cm,clip]{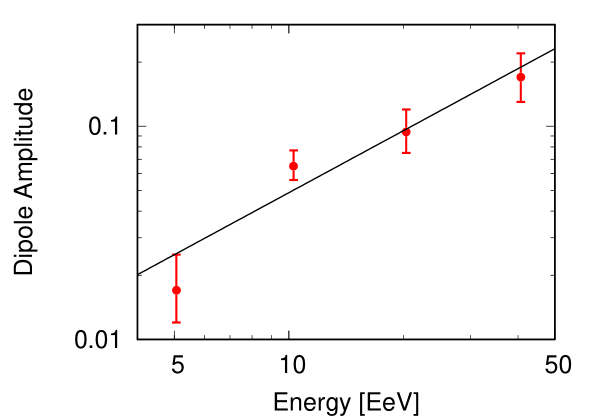}
\caption{Left: Directions of the 3D dipole for the four energy bins, in Galactic coordinates. Right: Evolution of the dipole amplitude (in units of the monopole) with the energy.}
\label{fig-2}
\end{figure}

In Fig.~\ref{fig-2}, the evolution with the energy of the dipole direction and amplitude is shown. The direction does not appear to significantly change with the energy, while the amplitude is growing. This
growth in amplitude could be due to the larger relative contribution from the nearby sources for increasing energies, whose distribution is more inhomogeneous \cite{Harari}, and/or to the growth of the mean rigidity, leading to smaller deflections and a larger dipolar amplitude \cite{Mollerach:2022aji}.

% \subsection{3D dipole and quadrupole above 4~EeV}
% \label{subsec-2}

% The analysis conducted in Sect.~\ref{subsec-1}
The analysis is repeated for the four energy bins, now including a quadrupolar component. It is observed that the quadrupolar components are not significant, and the dipole components are consistent with the results that we obtain when assuming only a dipole \cite{PierreAuger:2024fgl}.

\subsection{Modulation in R.A. down to 0.03~EeV}
\label{subsec-3}

\begin{figure}[h]
\centering
\includegraphics[width=5.2cm,clip]{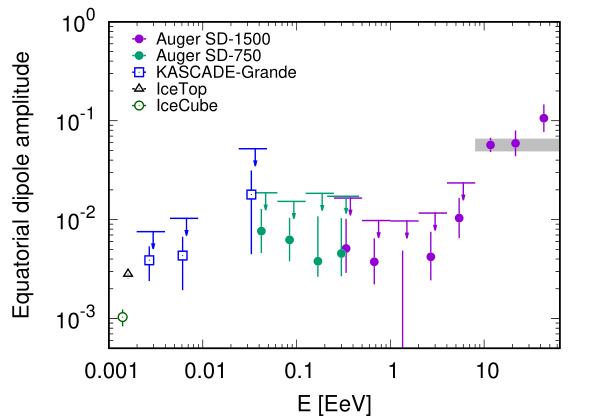}
\includegraphics[width=5.2cm,clip]{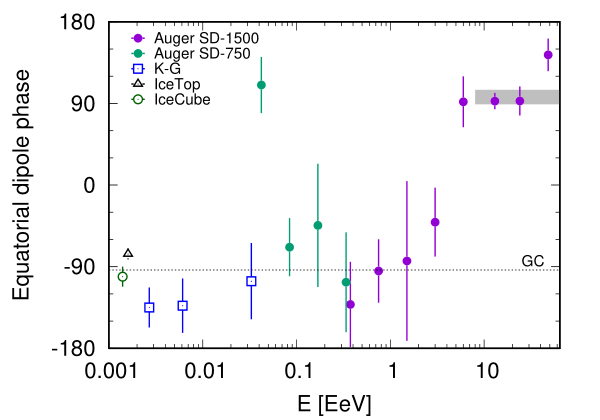}
\caption{Evolution of the equatorial dipole amplitude (Left) and phase (Right) with the energy. The results from the IceCube and KASCADE-Grande Collaborations are included at lower energies.}
\label{fig-3}
\end{figure}

For lower energies, only the equatorial component of the dipole was reconstructed, using the East-West method \cite{Lyberis:2011fga}.
In Fig.~\ref{fig-3}, the results for the equatorial dipole amplitude and phase evolution with the energy are shown. %, also including those obtained at lower energies by IceCube and KASCADE-Grande.
The equatorial dipole amplitude increases with the energy, from below 1\% to above 10\%, and the phase shifts from a direction close to the Galactic center to the opposite. This result appears to suggest a transition from a Galactic to an extragalactic origin of cosmic ray anisotropies around energies of a few EeV, though it is hard to imagine any Galactic mechanism able to accelerate cosmic rays up to such energies.

\subsection{Angular power spectrum above 4~EeV}
\label{subsec-4}

To search for anisotropies across various angular scales, it is convenient to decompose the distribution of observed events in spherical harmonics $Y_{\ell m}$. The harmonic coefficients $a_{\ell m}$ encode the variations on angular scales $\sim 180^{\circ}/ \ell$. The angular power spectrum \cite{Sommers:2000us} is defined as $C_{\ell} = \frac{1}{2\ell + 1} \sum_{m=-\ell}^{\ell} |a_{\ell m}|^2$.
Due to the incomplete sky coverage of the Observatory, a pseudo-power spectrum \cite{Deligny:2004dj} has to be computed ($\tilde{C}_{\ell}$), defined in terms of the pseudo-multipolar moments ($\tilde{a}_{\ell}$). The analysis is carried out in the same four energy bins as in Sect.~\ref{subsec-1}, and in the cumulative bin above 8~EeV. %The results are shown in Fig.~\ref{fig-4}.
Besides the significant dipolar pattern corresponding to $\tilde{C}_1$, in agreement with the Fourier analysis, the only $\tilde{C}_{\ell}$ values that stand above the 99\% CL of isotropic fluctuations are $\tilde{C}_{17}$ for the energy bin of 4-8~EeV and $\tilde{C}_8$ for the energy bin of 16-32~EeV. However, after a statistical penalization for searches over different multipoles and energy bins, their significance is 3.3\% and 26.5\% respectively. 

\section{Conclusion and outlook}
\label{sec-2}
In this contribution, we presented the results of large-scale anisotropy searches using UHECR arrival direction data collected at the Pierre Auger Observatory during 19 years of operation. The dipole was reconstructed via a separate Fourier analysis in R.A. and azimuth, with a significance of $6.8\sigma$ above 8~EeV, being well beyond the discovery level.
%Also, the evolution with the energy of the dipole direction and amplitude has been studied.
Its direction does not show any clear trend with the energy, while its amplitude is increasing, likely due to the larger relative contribution of nearby sources for increasing energies and/or by the growth of the mean rigidity.
The same analysis was conducted assuming that also a quadrupolar component is present, leading to non-significant results for the obtained quadrupolar moments. 
The dipole was studied also at lower energies, where only the reconstruction of the equatorial component was possible, using the East-West method. The results show an increasing trend of the equatorial dipole amplitude with the energy, while the phase is shifting from a direction close to Galactic center to the opposite direction, suggesting a possible transition from Galactic to extragalactic origins of UHECRs of increasing energies.
Finally, the angular power spectrum has been studied.
%across four energy bins and cumulatively above 8~EeV.
While the dipole component is always significant, except for the 4-8~EeV and the $\geq$ 32~EeV energy bins, in agreement with the 3D dipole analysis, only the $\tilde{C}_{17}$ and $\tilde{C}_8$ are above the 99\% CL of isotropic fluctuations for the energy bins of 4-8~EeV and 16-32~EeV respectively.
In general, the results align with model predictions, as detailed in \cite{PierreAuger:2024fgl}.
Future analyses using event-by-event mass-composition estimators thanks to AugerPrime \cite{AugerPrime} will enable further studies by separating the lighter and the heavier UHECR components.

% \begin{figure}[H]
% \centering
% \includegraphics[width=4.2cm,clip]{macro-latex-web-conf/figures/APS_4-8.png}
% \includegraphics[width=4.2cm,clip]{macro-latex-web-conf/figures/APS_8-16.png}
% \includegraphics[width=4.2cm,clip]{macro-latex-web-conf/figures/APS_16-32.png}
% \includegraphics[width=4.2cm,clip]{macro-latex-web-conf/figures/APS_Above32.png}
% \includegraphics[width=4.2cm,clip]{macro-latex-web-conf/figures/APS_Above8.png}
% \caption{Angular power spectrum measured in the four energy bins, as well as the cumulative energy bin above 8~EeV. The gray bands correspond to the 99\% CL from isotropic fluctuations, while the red lines represent the 99\% CL upper limits.}
% \label{fig-4}
% \end{figure}

% \begin{figure}[h]
% \centering
% \includegraphics[width=5cm,clip]{macro-latex-web-conf/figures/dipoleamplitudeHE2MRS.pdf}
% \includegraphics[width=5cm,clip]{macro-latex-web-conf/figures/quadrupoleamplitude2MRS.pdf}
% \caption{Please write your figure caption here}
% \label{fig-5}
% \end{figure}

\bibliography{biblio}
\end{document}